\documentclass[aps,prd,twocolumn,showpacs,superscriptaddress,groupedaddress,10pt]{revtex4-1}
\usepackage{graphicx,epstopdf,caption,subcaption}  
\usepackage{bm,braket,soul,color}        
\usepackage{amssymb,MnSymbol}   
\usepackage{hyperref}

\begin{document}

\title{Spatially extended Unruh-DeWitt detectors for relativistic quantum information}

\author{Antony R. Lee}
\author{Ivette Fuentes}\thanks{Previously known as Fuentes-Guridi and Fuentes-Schuller.}
\affiliation{School of Mathematical Sciences, University of Nottingham, University Park,
Nottingham NG7 2RD, United Kingdom}

\begin{abstract}
Unruh-DeWitt detectors interacting locally with a quantum field are systems under consideration for relativistic quantum information processing.  In most works, the detectors are assumed to be point-like and, therefore, couple with the same strength to all modes of the field spectrum. We propose the use of a more realistic detector model where the detector has a finite size conveniently tailored by a spatial profile. We design a spatial profile such that the detector, when inertial, naturally couples to a peaked distribution of Minkowski modes. In the uniformly accelerated case, the detector couples to a peaked distribution of Rindler modes. Such distributions are of special interest in the analysis of entanglement in non-inetial frames. We use our detector model to show the noise detected in the Minkowski vacuum and in single particle states is a function of the detector's acceleration.
\end{abstract}

\maketitle

\section{Introduction}
Relativistic quantum information is a multifaceted area that incorporates key principles from quantum field theory, quantum information and quantum optics to answer, primarily, questions of an information theoretic nature. In order to implement quantum information tasks in relativistic settings it is necessary to find suitable localised systems to store information. Moving point-like detectors coupled to quantum fields have been considered to carry quantum information in spacetime \cite{lin2008,lin2009,doukas2010}, perform teleportation \cite{lin2012}, and extract entanglement from the Minkowski vacuum \cite{reznik2003,steeg2009,lin2009}. For a review see \cite{alsing2012}. Our research program aims at developing new detector models which are more realistic and simpler to treat mathematically so they can be used in relativistic quantum information processing.

In this paper we utilise finite-size detectors \cite{takagi1986,grove1983,schlicht2004}, i.e. detectors with a position dependent coupling strength, which are not only more realistic but also have the advantage of coupling to peaked distributions of modes. We design Gaussian-type spatial profiles such that a uniformly accelerated detector naturally couples to peaked distributions of  Rindler modes. By expanding the field in terms of Unruh modes, we show the accelerated detector couples simultaneously to two peaked distribution of modes corresponding to left and right Unruh modes. As expected, the same detector interacts with a Gaussian distribution of Minkowski modes when it follows an inertial trajectory. In the Minkowski vacuum, the response of the detector has a thermal signature when it is uniformly accelerated and the temperature depends on the proper acceleration of the detector.

In the prototypical studies of quantum entanglement in non-inertial frames, observers are assumed to analyse states involving sharp frequency modes \cite{alsing2003,funentes-schuller2005}. In particular, recent works analysing the entanglement degradation between global modes seen by uniformly accelerated observers consider states of modes labelled by Unruh frequencies \cite{bruschi2010,bruschi2012,martinmartinez2010}. Our analysis provides further insight into the physical interpretation of the particle states which were analyzed in these works. Finite-size detectors are suitable to discuss such results from an operational perspective since sharp frequency modes are an idealization of the peaked distributions finite-size detectors couple to. We show the response of the finite-size detector when the state of the field has a single particle labelled by an Unruh frequency has a thermal term plus a noise term that depends on acceleration. Therefore, a degradation of global mode entanglement in non-inertial frames as a function of acceleration should be detected by uniformly accelerated finite-size detectors. Although global mode entanglement cannot be detected directly it can be extracted by Unruh-DeWitt type detectors becoming useful for quantum information tasks. Throughout our work we use natural units and the metric has signature $(-,+,+,+)$. 

\section{Detector Model}

The action of a detector interacting with a quantum field in the interaction picture is given by \cite{grove1983,takagi1986}, 
\begin{eqnarray}
\label{eqn:sysmtemaction}
S_{I}&=&\int dV\mathcal{M}\cdot\phi,
\end{eqnarray}
where $dV$ is the volume element for the spacetime and $\mathcal{M}$ is the monopole moment of the detector parametrised by arbitrary coordinates in a ($3+1$)-dimensional spacetime. We will assume a flat spacetime and the field $\phi$ to be a real massive scalar field which satisfies the Klein-Gordon equation,
\begin{eqnarray}
\nabla_{\mu}\nabla^{\mu}\phi-m^{2}\phi=0,
\end{eqnarray}
where $m\ge0$ is the free field mass and $\nabla_{\mu}$ denotes the covariant derivative on the spacetime. We require the detector to have, in a comoving reference frame with coordinates $(\tau,\vec{\zeta})$, a constant spatial profile. In this case the monopole moment of the detector factorizes into a temporal operator valued function and a spatial function $\mathcal{M}(\tau,\vec{\zeta})=M(\tau)f(\vec{\zeta})$~\cite{poisson2004,stephani2003,carroll2004}.  The temporal operator valued function describes the internal structure of the detector which we model by a two-level system of characteristic energy $\Delta$ therefore, $M(\tau):=\sigma_{-}e^{-i\tau\Delta}+\sigma_{+}e^{i\tau\Delta}$. The operators $\sigma_{+}$ and $\sigma_{-}$ create and annihilate, respectively, excitations in the internal structure of the detector. In this case, the action can be written in terms of the detector's proper time $\tau$,
\begin{eqnarray}
S_{I}&=&\int d\tau M(\tau)\tilde{\phi}(\tau),
\end{eqnarray}
where the field the detector couples to is given by,
\begin{eqnarray} 
\label{eq:Mfield}
\tilde{\phi}(\tau):=\int d^{3}\vec{\zeta}\sqrt{-g(\vec{\zeta)}}f(\vec{\zeta})\phi(\tau,\vec{\zeta}),
\end{eqnarray}
and $g=\text{det}(g_{\mu\nu})$ is the determinant of the metric tensor.  We assume the center of the detector follows a classical trajectory in spacetime and the spatial profile $f(\vec{\zeta})$ determines how the detector couples to the field along the trajectory. This function, which must be real to ensure the action is Hermitian, can be interpreted as a position dependent coupling strength. Consider $u_{\vec{\nu}}(\vec{\zeta}(\tau))$ to be field solutions to the Klein-Gordon equation evaluated along a point-like worldline parametrised by $\tau$ corresponding to the center point of the detector. The frequencies $\vec{\nu}$ of the modes are determined by an observer comoving with the center of the detector. The Hamiltonian in terms of these modes takes the form,
\begin{eqnarray}
\label{hamiltonian}
H_{I}&=&M(\tau)\cdot\int d^3\vec{\nu} \tilde{f}(\vec{\nu})\left(u_{\vec{\nu}}(\tau)a_{\vec{\nu}}+h.c.\right),
\end{eqnarray}
where $a^{\dagger}_{\vec{\nu}}$ and $a_{\vec{\nu}}$ are creation and annihilation operators associated to the field modes of frequency $\vec{\nu}$. The frequency distribution $\tilde{f}(\vec{\nu})$ corresponds to a transformation of  $f(\vec{\zeta})$ into frequency space. In the ideal case where the detector is considered to be point-like, the spatial profile is $f(\vec{\zeta})=\delta^{3}(\vec{\zeta}-\vec{\zeta}')$ (here $ \delta^{3}(\vec{\zeta}-\vec{\zeta}'):=\delta(\zeta_{1}-\zeta_{1}')\delta(\zeta_{2}-\zeta_{2}')\delta(\zeta_{3}-\zeta_{3}')$ is the three dimensional Dirac delta distribution). The detector couples locally to the field and the coupling strength is uniformly equal for all frequency modes. When we model a finite-size detector, which corresponds to a more realistic situation, the detector couples naturally to a distribution of field modes. The frequency distribution will be determined by the spatial profile. In this sense the field $\tilde{\phi}(\tau)$ corresponds to a {\it window} of frequencies. 

\section{Inertial Trajectory}
Now we specify a trajectory for the detector.  When the center of the detector follows an inertial trajectory it is convenient to use Minkowski coordinates $(t,\vec{x})$ where  $\vec{x}:=({x},{y},{z})$. In this case, the proper time of a comoving observer is $\tau=t$ and we can also write the comoving spatial coordinates as $\vec{\zeta}=\vec{x}$. The solutions to the Klein-Gordon equation correspond to plane waves,
\begin{eqnarray}
u_{\vec{k}}(t,\vec{x})&:=&\frac{1}{\sqrt{2(2\pi)^{3}\omega}}e^{-i\omega t+i\vec{k}\cdot\vec{x}},
\end{eqnarray}
where the frequency of the mode $\omega\equiv\sqrt{\vec{k}\cdot\vec{k}+m^{2}}$ is strictly positive and $\vec{k}\in\mathbb{R}^{3}$ denotes the momentum  $\vec{k}:=(k_x,k_y,k_z)$. The inner product satisfies $\left(u_{\vec{k}},u_{\vec{k}'}\right)=\delta^{3}(\vec{k}-\vec{k}')$ \cite{crispino2008}. In this case the creation and annihilation operators associated to the Minkowski field modes satisfy $\left[a_{\vec{k}},a^{\dag}_{\vec{k}'}\right]=\delta^{3}(\vec{k}-\vec{k}')$. The field can be expanded in Minkowski modes as
\begin{equation}
\phi(t,\vec{x})=\int d^{3}\vec{k}\left(a_{\vec{k}}u_{\vec{k}}(t,\vec{x})+\text{h.c.}\right).
\end{equation}
From this, the frequency distribution expressed in Minkowski modes is, 
\begin{eqnarray}
\label{eqn:fourier_transformation}
\tilde{f}(\vec{k})&=&\int d^{3}\vec{x} f(\vec{x})e^{+i\vec{k}\cdot\vec{x}},
\end{eqnarray}
which is the Fourier transform of the spatial profile $f(\vec{x})$. 

We now design a spatial profile tailored so that the corresponding frequency detection window of the detector is a Gaussian distribution of modes peaked around a Minkowski frequency $\vec{\lambda}$. This choice is motivated by early works on relativistic entanglement
where the states analysed involved sharp frequencies $\Omega$ and $\Omega'$. For example, the Bell state,
\begin{equation}
|\phi\rangle=\frac{1}{\sqrt{2}}(|0\rangle_{\Omega}|0\rangle_{\Omega'}+|1\rangle_{\Omega}|1\rangle_{\Omega'}),
\end{equation}
which was analysed in a flat (1+1)-dimensional space in the uncharged massless bosonic~\cite{bruschi2010} and charged~\cite{bruschi2012} case. Entanglement for Bell states in non-inertial frames was also discussed for Dirac fields \cite{martinmartinez2009,martinmartinez2010,martinmartinez2012-2}. 
Sharp frequency states, $|1\rangle_{\Omega}=a_{\Omega}^{\dag}|0\rangle$, are an idealisation of Gaussian wave-packets of the form
\begin{equation}
\label{Eq:particlesmearing}
|1_{\lambda}\rangle=\int dk \Phi(\lambda,k)a_{k}^{\dag}|0\rangle,
\end{equation}
where $\Phi(\lambda,k)$ is a Gaussian distribution centred around $\lambda$ \cite{bruschi2010}. Our detector model will be useful to investigate questions of entanglement in non-inertial frames from an operational perspective and extract entanglement for relativistic quantum information processing. An interesting question, which we intend to address in following work, is how much entanglement can be extracted by our detectors from field mode entangled states as a function of the detector's acceleration.
\begin{figure}[t]
\includegraphics[width=\linewidth]{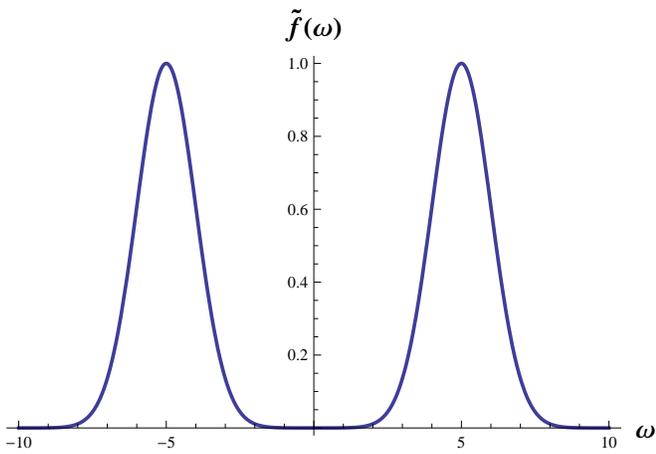}
\caption{\label{gaussianPeakingfig} $(1+1)$ dimensional example of a frequency distribution peaked around $\pm\lambda$ given by Eq.(\ref{gaussianPeaking}) for $\sigma=1$ and $\lambda=5$. This frequency distribution peaks around around the desired frequency $\lambda$ but has a double peaking due to the two exponential terms. In the $(1+1)$ massless case, the field is expanded as an integral over $\omega>0$ and so the peak in the $\omega<0$ region does not contribute.} 
\end{figure}
\begin{figure}[b]
\includegraphics[width=\linewidth]{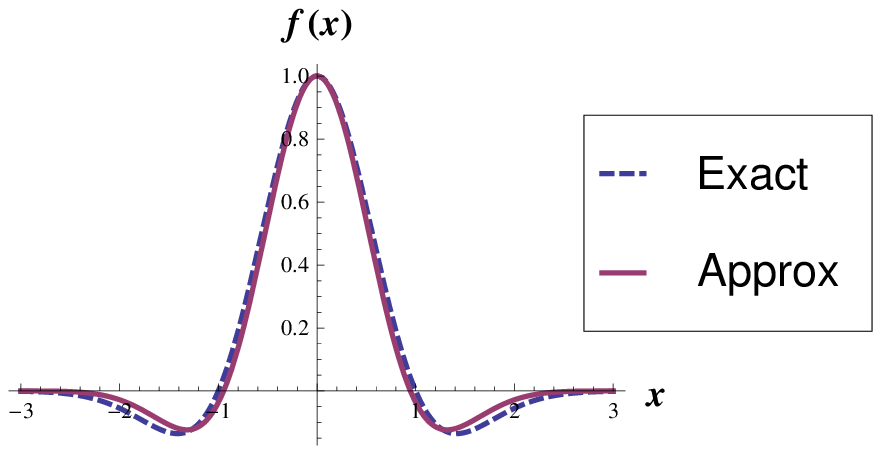}
\caption{\label{fig:0_1_coupling} Comparison of exact coupling strength (as modelled with a quantum harmonic oscillator) and our approximate coupling strength. The choices are: quantum numbers $(n,m)=(0,1)$ and parameters $(\lambda,\sigma)=(1.66,1/\sqrt{0.89})$.}
\end{figure}
\begin{figure}[b]
\includegraphics[width=\linewidth]{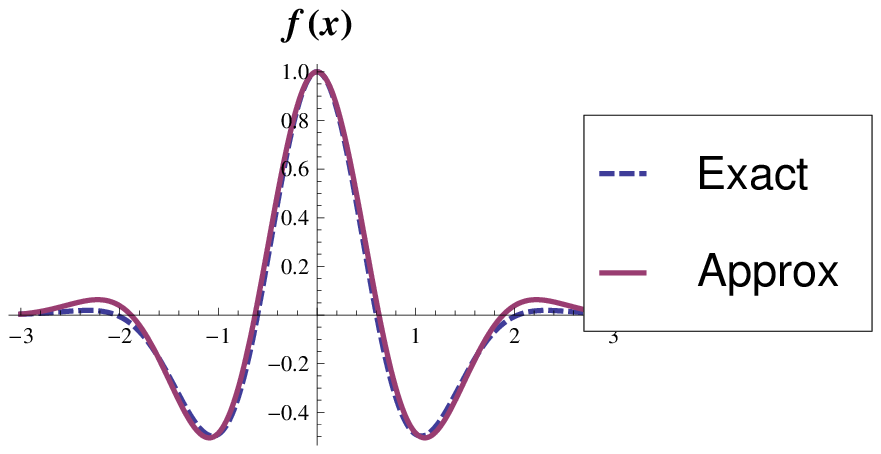}
\caption{\label{fig:0_3_coupling} Comparison of exact coupling strength (as modelled with a quantum harmonic oscillator) and our approximate coupling strength. The choices are: quantum numbers $(n,m)=(0,3)$ and parameters $(\lambda,\sigma)=(2.5,1)$.}
\end{figure}
\begin{figure}[b]
\includegraphics[width=\linewidth]{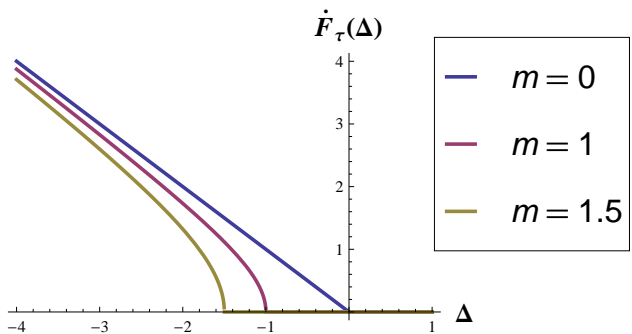}
\caption{\label{fig:3+1inertialdeltaeps} The transition rate for an inertial, Dirac delta spatial profile (i.e. point-like) detector probing a $(3+1)$ massive field. We have plotted the transition rate for mass values $m=0,1,1.5$. We observe that if the detector is in its excited state ($-\Delta>0$), its energy gap needs to be larger than the mass of the field to undergo spontaneous emission. Also, for all energy gaps $-\Delta<m$, the detector cannot undergo spontaneous emission i.e. it is stable in its excited state.}
\end{figure}
We find that a Gaussian frequency window of width $\sigma$ centred around frequency $\vec{\lambda}$ as shown in FIG.(\ref{gaussianPeakingfig}), can be engineered by choosing the following spatial profile,
\begin{eqnarray}
\label{gaussianprof}
f(\vec{x})&=&n_{\sigma}e^{-\frac{1}{2}\sigma^{-2}\vec{x}\cdot\vec{x}}\left(e^{-i\vec{\lambda}\cdot\vec{x}}+e^{+i\vec{\lambda}\cdot\vec{x}}\right),
\end{eqnarray}
which corresponds to a Gaussian distribution multiplied by a superposition of planes waves of opposite momentum $\vec{\lambda}$. $n_{\sigma}$ is a normalisation constant. This spatial profile is of great mathematical convenience and, moreover, similar couplings naturally arise in the interaction of harmonic oscillators and quantum fields. For example, consider the coupling between the two-level system and the field is given by~\cite{walls2008,martinmartinez2012},
\begin{eqnarray}
\label{eqn:derived_coupling_strength}
f(x)=\varphi_{-}^{*}(x)\partial_{x}\varphi_{+}(x),
\end{eqnarray}
where $\varphi_{-}(x)$ and $\varphi_{+}(x)$ are the ground and excited wavefuntions of the detector. One can consider the two energy levels to be two eigenstates of a quantum harmonic oscillator. In this case, the interaction strength will be an oscillatory Gaussian-type function that can be approximated by the coupling strength we propose in Eq.~(\ref{gaussianprof}). 
The wavefunctions of the two-level system in this case can be taken to be Hermite functions of the form,
\begin{subequations}
\begin{align}
\varphi_{n}\,&=\,N_{n}e^{-\frac{1}{2}x^{2}}\mathrm{H}_{n}(x),\\
\varphi_{m}\,&=\,N_{m}e^{-\frac{1}{2}x^{2}}\mathrm{H}_{m}(x),
\end{align}
\end{subequations}
where $N_{k}$ are normalisation constants, $\mathrm{H}_{k}$ are Hermite poynomials and we impose $n<m$. Inserting these wavefunctions into the coupling strength Eq.~(\ref{eqn:derived_coupling_strength}), and using the well-known recursion relations of the Hermite functions, we find,
\begin{equation}
f(x)\,=\,\varphi_{n}^{*}\,\left(\sqrt{\frac{m}{2}}\varphi_{m-1}-\sqrt{\frac{m+1}{2}}\varphi_{m+1}\right).
\end{equation}
From the properties of the Hermite polynomials, we know that the quantum numbers of the quantum harmonic oscillator wavefunctions need to take particular values. To be precise, we need $n=2k$ (an even integer) and $m=2p+1$ (an odd integer). This is to ensure our example coupling strength accurately approximates the quantum harmonic oscillator coupling. A different choice of pairing would either result in a coupling strength that is overall an odd function or would necessarily be zero at the origin, both contradicting the form of Eq.~(\ref{gaussianprof}). Assuming the ground state to be given by $n=0$, it can be easily shown that the pairs $(n,m)=(0,1)$ and $(0,3)$ can be accurately approximated by our coupling strength. The two pairs correspond to a bound state between the lowest energy eigenstate of the quantum harmonic oscillator and it's first and third energy states. As an illustrative example, we consider the system which is limited to interactions between the ground and 3\textsuperscript{rd} excited state. In this case the normalised wave functions are,
\begin{subequations}
\begin{align}
\varphi_{0}\,=\,N_{0}e^{-\frac{1}{2}x^{2}}\mathrm{H}_{0}(x),\\
\varphi_{3}\,=\,N_{3}e^{-\frac{1}{2}x^{2}}\mathrm{H}_{3}(x).
\end{align}
\end{subequations}
Consequently, it can be seen that the corresponding coupling strength, $F(x)\sim(2x^4-9x^2+3)e^{-x^{2}}$ , is approximated by a coupling of the form~(\ref{gaussianprof}) with the parameter choices $(\lambda,\sigma)=(2.5,1)$. Our coupling strength would therefore be a close approximation for a bound quantum harmonic oscillator. See Figs.~(\ref{fig:0_1_coupling}, \ref{fig:0_3_coupling}) for plots showing the $(n,m)=(0,1)$ and $(0,3)$ systems respectively. We bring to attention reference~\cite{colombe2007} where a two dimensional version of the coupling function~(\ref{gaussianprof}) has been experimentally demonstrated in a BEC-cavity system. It should be noted, however, if the bound system is between high energy level wavefunctions then our assumed coupling strength would no longer be a good approximation to the physical coupling strength. This is due to the asymptotic form of the harmonic oscillator wavefunctions which approach either a purely cosine or sine form and lose their Gaussian nature. As a final comment, we would like to point out that a physically realised spatial profile will generally be given by an effective coupling of the internal degrees of freedom of the detector and the field. In practice, it will have contributions from more than just the wavefunction of the detector.
\begin{figure}[b]
\includegraphics[width=\linewidth]{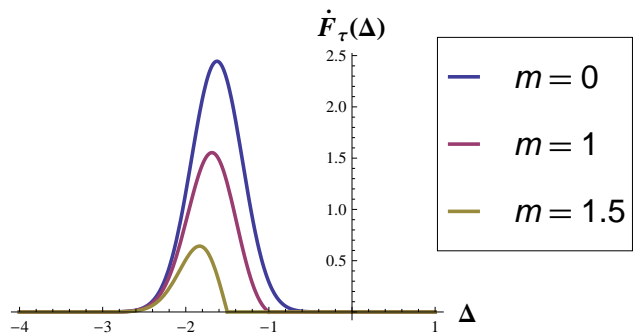}
\caption{\label{fig:3+1inertialgaussian} The transition rate for an inertial, double peaked Gaussian spatial profile detector (see Eq.(\ref{gaussianprof})) probing a $(3+1)$ massive field. We have plotted the transition rate for mass values $m=0,1,1.5$. We observe that if the detector is in its excited state ($-\Delta>0$), for high energy gaps the detector's transition rate is negligibly small. This implies that for large energy gaps, a non-point-like detector is stable i.e. it will not undergo spontaneous emission.}
\end{figure}
To continue, using Eq.~(\ref{eqn:fourier_transformation}), the spatial profile is transformed into the momentum distribution,
\begin{eqnarray}
\label{gaussianPeaking}
\tilde{f}(\vec{k})=e^{-\frac{1}{2}\sigma^{2}(\vec{k}-\vec{\lambda})\cdot(\vec{k}-\vec{\lambda})}+e^{-\frac{1}{2}\sigma^{2}(\vec{k}+\vec{\lambda})\cdot(\vec{k}+\vec{\lambda})}.
\end{eqnarray}
This means that in order to couple our detector to a peaked Gaussian distribution of modes centered around $\vec{\lambda}$ it is necessary to engineer a field-detector coupling strength which not only is peaked around the atom's trajectory but also oscillates with position. Sharp frequency modes $\tilde{f}(\vec{k})=\delta^{3}(\vec{k}-\vec{\lambda})+\delta^{3}(\vec{k}+\vec{\lambda})$ are obtained when  $f(\vec{x})\sim\text{exp}(-i\vec{\lambda}\cdot\vec{x})+\text{exp}(+i\vec{\lambda}\cdot\vec{x})$. In the massless $(1+1)$-dimensional case the frequency distribution obtained from a given spatial profile is defined as a function of $\omega\ge 0$ only. Therefore, given the window profile peaks are sufficiently narrow and separated, the second peaking corresponding to the $\omega<0$ region does not contribute to the frequency window in this case. The field to which the detector couples, given by Eq.(\ref{eq:Mfield}), is therefore,
\begin{eqnarray}
\phi(\tau)=\int d\omega N_{\omega}e^{-\frac{1}{2} \sigma^{2}({\lambda} -\omega )^2}\left[e^{-i\omega t}a_{\omega}+e^{+i\omega t}a^{\dagger}_{\omega}\right].
\end{eqnarray}
In the general case, the frequency window is peaked around two modes corresponding to negative and positive momentum. Also see \cite{martinmartinez2012} which, complementary to this work, looks in detail at the coupling of Unruh-DeWitt detectors to Minkwoski modes. Eq.(\ref{gaussianPeaking}) shows there is a trade-off between the width of the frequency window and the spatial profile of the detector.
 
We now proceed to calculate the instantaneous transition rate of the detector model we have introduced. The transition rate is a function of the detector's energy gap $\Delta$ given by \cite{schlicht2004}
\begin{eqnarray}
\label{transitionrate}
\dot{F}_{\tau}(\Delta)&:=&2\int_{0}^{\infty}ds Re\left[e^{-i\Delta s}W(\tau,\tau-s)\right],
\end{eqnarray}
where $W(\tau,\tau'):=\bra{\psi}\phi(\tau)\phi(\tau')\ket{\psi}$ is the so-called Wightman-function and $\ket{\psi}$ denotes the state of the field.  Note we have assumed that the detector is turned on in the distant past. Expanding the field in terms of Minkowski modes we find that the vacuum transition rate for a stationary detector is
\begin{eqnarray}
\label{eqn:inertialtransitionrate}
\dot{F}_{\tau}(\Delta)=\Theta(-\Delta-m)\Xi(\Delta),
\end{eqnarray}
where 
\begin{eqnarray}
\label{eqn:xiflat}
\Xi(\Delta)=\sqrt{(-\Delta)^{2}-m^{2}}|\tilde{f}(-\Delta)|^{2},
\end{eqnarray}
and $\Theta(x)$ is the Heavisde theta function defined as 
\begin{eqnarray}
\Theta(x)&=&\left\{\begin{array}{ccl} 0 & : & x<0\\ 1 & : & x\ge 0\end{array}\right. .
\end{eqnarray}
We have explicitly written $-\Delta$ in Eq.~(\ref{eqn:xiflat}) to emphasise that in the limit $m\rightarrow 0$ we recover the standard literature result. Note that in the above result we have explicitly assumed $\tilde{f}(\vec{k})=\tilde{f}(|\vec{k}|)$ i.e. the Fourier transform of the spatial profile $f(\vec{x})$ depends on the magnitude of $\vec{k}$ only. The result is explicitly independent of the time parameter $\tau$. This can be traced back to the fact that an inertial trajectory is a stationary orbit in flat spacetime. The interesting result here is that the transition rate of the detector is tempered by the square of the frequency distribution $\tilde{f}$. As examples, we have plotted the transition rate Eq.~(\ref{eqn:inertialtransitionrate}) for a Dirac delta peaked profile ($f(x)\sim\delta(x)$) and a double peaked Gaussian profile Eq.~(\ref{gaussianPeaking}) in FIG.~(\ref{fig:3+1inertialdeltaeps})~and FIG.~(\ref{fig:3+1inertialgaussian}) respectively. In FIG.~(\ref{fig:3+1inertialdeltaeps}), we can see that point-like excited detectors always have a possibility of undergoing spontaneous emission for $-\Delta>m$. Outside of this region, the detector will either remain excited or in its ground state. In other words, spontaneous emission (or absorption) is not possible. In FIG.~(\ref{fig:3+1inertialgaussian}), we observe how a Gaussian type spatial profile modifies the transition rate of an inertial detector. In stark contrast to the point-like detector, for a detector in its excited state with a large energy gap, the transition rate is negligible outside a ``resonance" region. These regions effectively tell us the detector is sensitive to modes with energy $\sim|\Delta|$.

We are also interested in analysing the response of the detector when the field has a single Minkowski excitation, \begin{eqnarray}
\ket{1_{\Phi}}&:=&\int d^{3}\vec{k} \Phi(\vec{k})a_{\vec{k}}^{\dag}\ket{0},
\end{eqnarray}
where we define a delta normalised state to have the property $\int d^{3}\vec{k} |\Phi(\vec{k})|^{2}=\delta^{3}(0)$ and a properly normalised state to have the property $\int d^{3}\vec{k} |\Phi(\vec{k})|^{2}=1$. This state is the generalization of Eq.(\ref{Eq:particlesmearing}) to three spatial dimensions. The Wightman-function in this case is \begin{eqnarray}
W(\tau,\tau')&=& \bra{1_{\Phi}}\phi(\tau)\phi(\tau')\ket{1_{\Phi}}.
\end{eqnarray}
Writing the states and the field in terms of the Minkowski modes and normal ordering the associated operators we find that
\begin{equation}
\begin{aligned}
\label{wightmann1p}
W(\tau,\tau')&=&\hspace{-3mm}\int d^{3}\vec{k} |\Phi(\vec{k})|^{2}\cdot \bra{0}\phi(\tau)\phi(\tau')\ket{0}\\
& &+2\text{Re}\left[I_{\Phi}^{*}(\tau)I_{\Phi}(\tau')\right],
\end{aligned}
\end{equation}
where  
\begin{eqnarray}
I_{\Phi}(\tau)=\int d^{3}\vec{k}\Phi(\vec{k})\tilde{f}(\vec{k})\frac{1}{\sqrt{\omega}}e^{-i\omega\tau}.
\end{eqnarray}  
We notice there are two terms in the Wightman-function. The first one corresponds to the vacuum state and the second is the contribution from the particle present in the field. 

The single particle contribution factorizes into two independent functions of $\tau$ and $\tau'$. This allows us to analyse the transition rate with relative ease. Substituting Eq.~(\ref{wightmann1p}) into Eq.~(\ref{transitionrate}) we obtain an expression which depends on
\begin{eqnarray}
\label{iota}
\iota_{\tau}(\Delta)&:=&\int_{0}^{\infty} ds e^{-is\Delta}I_{\Phi}(\tau-s).
\end{eqnarray}
This integral is essentially a Fourier transform of $I_{\Phi}(\tau-s)$  in the $s$ variable and can be computed, either analytically or numerically, for specifically chosen $\tilde{f}$ and $g$.  Employing the Riemann-Lebesgue lemma, which can only be used for functions which are integrable on the real line, one shows that $I_{\Phi}(\tau)\rightarrow 0$ as $\tau\rightarrow\pm\infty$ as long as $\tilde{f}$ and $g$ are well-behaved. $I_{\Phi}(\tau)$ vanishes in the distant past and future where the detector is responding only to vacuum fluctuations. In other words, the detector only observes a constant spectrum in these asymptomatic regions. In the intermediate regions the oscillatory response is due to the presence of the particle.

\section{Accelerated Trajectory}
We now consider a detector following a uniformly accelerated trajectory.  Conformally flat Rinder coordinates $\underline{\xi}=(\tau,\vec{\xi})=(\rho,\xi,y,z)$ are a convenient choice in this case.  The transformation between Rindler and Minkowski coordinates is given by
\begin{eqnarray}
\label{eq:rindler}
\begin{array}{ccl} t & = & a^{-1}e^{a\xi}\sinh(a\rho)\\ x & = & a^{-1}e^{a\xi}\cosh(a\rho)\end{array},
\end{eqnarray}
where $a$ is a postive parameter and the other spatial coordinates do not change i.e. $y=y$, $z=z$. This transformation holds for the spacetime region $|t|>x$ which is called the right Rindler wedge. The coordinate transformation for $|t|>-x$ (left region) differs from Eq.~\eqref{eq:rindler} only by an overall sign in the $x$ and $t$ coordinates. The coordinates are tailored specifically to the trajectory $\vec{\xi}=\vec{0}$ so that an observer travelling along this worldline will measure a proper acceleration $\sqrt{-A^{\mu}A_{\mu}}=a$ and the proper time is parametrised by the coordinate time $\rho$.
The Klein-Gordon equation for a massive bosonic field in a (3+1)-dimensional flat spacetime in this case takes the form
\begin{eqnarray}
\partial_{\rho\rho}\phi-\left[\partial_{\xi\xi}+e^{2a\xi}(\partial_{yy}+\partial_{zz})-m^{2}e^{2a\xi}\right]\phi=0 ,
\end{eqnarray}
and the solutions are the Rindler modes \cite{bruschi2010,crispino2008}
\begin{equation}
\begin{aligned}
u_{\Omega,\vec{k}_{\perp},\alpha}(\rho,\vec{\xi})&:= N_{\Omega/a}K_{i\Omega/a}\left(Ma^{-1}e^{a\xi}\right)e^{-i\Lambda_{\alpha}(\rho,\vec{\xi})}\\
\Lambda_{\alpha}(\rho,\vec{\xi})&:=\alpha \rho \Omega-\vec{k}_{\perp}\cdot\vec{x}_{\perp},
\end{aligned}
\end{equation}
\begin{figure}[b]
\includegraphics[width=\linewidth]{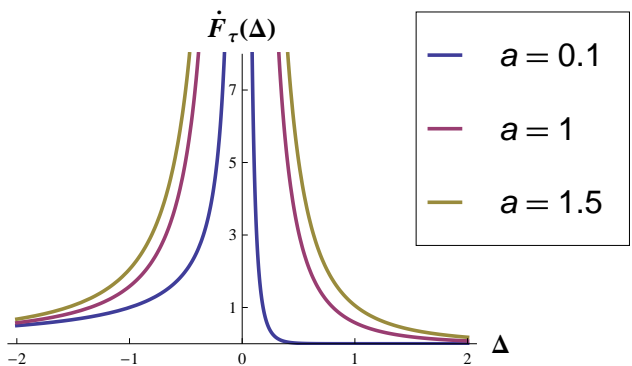}
\caption{\label{fig:1+1accelerateddelta} The transition rate for an accelerated, Dirac delta spatial profile detector probing a $(1+1)$ massless field. We have plotted the transition rate for acceleration values $a=0.1,1,1.5$. We see that the transition rate is zero only for asymptotically large values of $\Delta$. The divergent region $\Delta=0$ is due to the nature of the massless $(1+1)$ field.}
\end{figure}
with $M=\sqrt{\vec{k}_{\perp}\cdot\vec{k}_{\perp}+m^{2}}$ and  $N_{\Omega/a}$ is the mode normalisation constant. The functions $K_{i\Omega/a}(R)$ are modified Bessel functions of the second kind. Here $\vec{x}_{\perp}:=(y,z)$ and $\vec{k}_{\perp}:=(k_{y},k_{z})$ are position and momentum vectors perpendicular to the direction of acceleration. $\Omega$ is strictly positive and denotes the Rindler frequency and  $\alpha=+1(-1)$ corresponds to right (left) Rindler regions,  respectively. The canonical orthonormality relation for the $(3+1)$ massive field is,
\begin{equation}(u_{\Omega,\vec{k}_{\perp},\alpha},u_{\Omega',\vec{k}^{\prime}_{\perp},\alpha'})=\delta(\Omega-\Omega')\delta^{2}(\vec{k}_{\perp}-\vec{k}_{\perp}')\delta_{\alpha\alpha'},
\end{equation}and commutation relations satisfy
\begin{equation}[a_{\Omega,\vec{k}_{\perp},\alpha},a^{\dag}_{\Omega'\vec{k}^{\prime}_{\perp},\alpha'}]=\delta(\Omega-\Omega')\delta^{2}(\vec{k}_{\perp}-\vec{k}_{\perp}')\delta_{\alpha\alpha'}.
\end{equation} 
From our coordinate definitions Eq.~\eqref{eq:rindler}, we choose the detector to be travelling in the right Rindler wedge. This implies that our comoving coordinates can be parametrised as $\tau=\rho$ and $\vec{\zeta}=\vec{\xi}$. The field expansion in terms of the parametrised Rindler modes is
\begin{figure}[b]
\includegraphics[width=\linewidth]{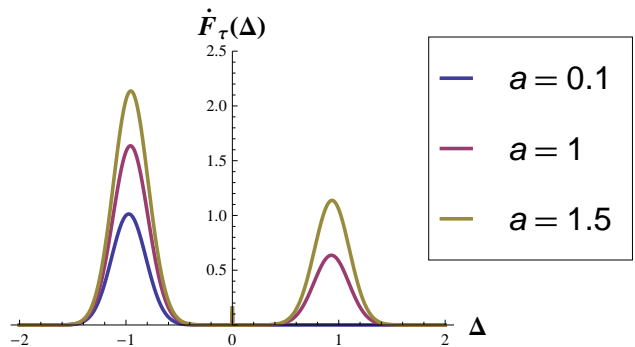}
\caption{\label{fig:1+1acceleratedgaussianeps} The transition rate for an accelerated, double peaked Gaussian spatial profile detector probing a $(1+1)$ massless field. We have plotted the transition rate for acceleration values $a=0.1,1,1.5$. We see that there are resonant values for $\Delta$ where probability of the detector undergoing spontaneous emission (or absorption) is increased. This can be traced back to the Gaussian spatial profile which causes the detector to interact more strongly with modes with energy $\sim|\Delta|$.}
\end{figure}
\begin{equation}
\label{eq:rinderexpansion}
\phi(\tau,\vec{\xi})=\int d\Omega d^{2}\vec{k}_{\perp}\left[u_{\Omega,\vec{k}_{\perp},+}(\rho,\vec{\xi})a_{\Omega,\vec{k}_{\perp},+}+\text{h.c.}\right].
\end{equation}
Note the left Rindler modes do not appear in~Eq.(\ref{eq:rinderexpansion}) as the detector is assumed to be moving in the right Rindler wedge. The explicit form of the accelerated detectors frequency distribution is
\begin{eqnarray}
\label{Eq:acceleratedsmearingexpression}
\tilde{f}(\Omega,\vec{k}_{\perp})=\!\!\int\!\! d^{3}\vec{\xi} e^{2a\xi}f(\vec{\xi})K_{i\Omega/a}\left(Ma^{-1}e^{a\xi}\right)e^{+i\vec{k}_{\perp}\cdot\vec{x}_{\perp}}
\end{eqnarray}
The most significant difference between the inertial and the accelerated frequency distributions is the appearance of a non-trivial metric factor and the Bessel function. Note also that for both massless and massive fields, the Rindler modes are defined as an integral over $\Omega\in\mathbb{R}^{+}$, unlike the Minkwoski mode case. Eq.(\ref{Eq:acceleratedsmearingexpression}) is a Fourier transform in the $y$ and $z$ coordinates, however, it is a non-standard integral transformation in the $\xi$ coordinate. Reminiscent of a Hankel transformation, we expect our desired properties of arbitrary mode peaking to still hold. Using the integral representation of the modified Bessel function for the second kind
\begin{eqnarray}
\!\! K_{i\Omega/a}(R)=\frac{\sqrt{\pi}\left(\frac{1}{2} R\right)^{i\Omega/a}}{\Gamma(i\Omega/a +1/2)}\!\int_{0}^{\infty}\!\!\!\! dt\frac{\left(\sinh(t)\right)^{2i\Omega/a +1}}{e^{R\cosh(t)}},
\end{eqnarray}
valid for $\Omega/a>0$ and $R>0$, we can write the frequency distribution as a Fourier type integral that takes the form
\begin{eqnarray}
\tilde{f}(\vec{k})=\int d^{3}\vec{\xi} \beta(\vec{\xi})e^{i \vec{\xi}\cdot\vec{k}},
\end{eqnarray}
where now $\vec{k}=(\Omega+\delta,k_{y},k_{z})$ and
\begin{eqnarray}
\beta(\vec{\xi})=\frac{\sqrt{\pi}\left(\frac{1}{2}\frac{M}{a}\right)^{i\Omega/a}}{\Gamma(i\Omega/a +1/2)}f(\vec{\xi})\int_{0}^{\infty}\!\!\!dr \frac{(\sinh(r))^{2i\Omega/a}}{e^{\frac{M}{a}e^{a\xi}\cosh(r)}},
\end{eqnarray}
$\delta$ is a phase that is acquired from the integral representation of the modified Bessel function. This shows that, in principle, the standard properties of the Fourier transformation can be used to design a detector profile such that we obtain a peaked distribution in momentum space. For a concrete example, we shall consider the massless $(1+1)$ field case. The appropriate transformation, in terms of Rindler modes, is given by
\begin{eqnarray}
\label{1plus1fourier}
\tilde{f}(\Omega)=\int d\xi e^{2a\xi} f(\xi)e^{i\Omega\xi},
\end{eqnarray}
and the spatial profile we propose in this case is 
\begin{eqnarray}
\label{Eq:modifedspatialprofile}
f(\xi)&=&N(\sigma)e^{-2a\xi}e^{-\frac{1}{2}\sigma^{-2}{\xi^2}}\left(e^{-i{\tilde{\lambda}}{\xi}}+e^{+i{\tilde{\lambda}}{\xi}}\right),
\end{eqnarray}
which includes the conformal metric factor that arises from the Rindler coordinate transformation. Here $N({\sigma})$ is a normalization constant and $\tilde{\lambda}$ dictates a preferred mode frequency. This profile reduces the integral transformation~Eq.(\ref{1plus1fourier}) to a standard Fourier transformation and the resulting frequency distribution is
\begin{eqnarray}
\tilde{f}(\Omega)=e^{-\frac{1}{2} \sigma^{2}(\tilde{\lambda} -\Omega )^2}+e^{-\frac{1}{2} \sigma ^2(\tilde{\lambda} +\Omega )^2}.
\end{eqnarray}
Substituting this frequency distribution into  Eq.(\ref{eq:Mfield}), we obtain
\begin{eqnarray}
\phi(\tau)=\int d\Omega N_{\Omega/a}e^{-\frac{1}{2} \sigma^{2}(\tilde{\lambda} -\Omega )^2}\left[e^{-i\Omega\tau}a_{\Omega,I}+h.c.\right].
\end{eqnarray}
Therefore, our detector couples to a Gaussian distribution centered around a Rindler frequency $\tilde{\lambda}$. In the limiting case where the acceleration goes to zero, the Rindler frequency goes to the Mickowski frequency i.e. $\Omega\rightarrow\omega$, and the spatial profile reduces to the Minkwoski profile given by Eq.({\ref{gaussianprof}}).  

We would now like to expand the field in terms of Unruh modes which pay an important role in the literature. These modes are given by \cite{bruschi2010,crispino2008},
\begin{eqnarray}
\begin{array}{ccl} 
\!\! u_{\Omega,\vec{k}_{\perp},R}&=&\cosh(r_{\Omega/a})u_{\Omega,\vec{k}_{\perp},+}+\sinh(r_{\Omega/a})u_{\Omega,\vec{k}_{\perp},-}^{*},\\
\!\! u_{\Omega,\vec{k}_{\perp},L}&=&\cosh(r_{\Omega/a})u_{\Omega,\vec{k}_{\perp},-}+\sinh(r_{\Omega/a})u_{\Omega,\vec{k}_{\perp},+}^{*},
\end{array}
\end{eqnarray}
where $\tanh(r_{\Omega/a})=e^{-\pi\Omega/a}$. Upon parametrising the modes with our accelerated comoving coordinates, i.e. $(\rho,\vec{\xi})=(\tau,\vec{\xi})$, and noting that the left Rindler modes have no support in the right Rindler wedge, we find the Unruh modes reduce to
\begin{eqnarray}
\begin{array}{ccl} 
u_{\Omega,\vec{k}_{\perp},R}(\tau,\vec{\xi})&=&\cosh(r_{\Omega/a})u_{\Omega,\vec{k}_{\perp},+}(\tau,\vec{\xi}),\\
u_{\Omega,\vec{k}_{\perp},L}(\tau,\vec{\xi})&=&\sinh(r_{\Omega/a})u_{\Omega,\vec{k}_{\perp},+}^{*}(\tau,\vec{\xi}).
\end{array}
\end{eqnarray}
In the case the field is massless and in $(1+1)$-dimensions, the detector interacts with two peaked distributions corresponding to right and left Unruh modes respectively,
\begin{eqnarray}
\phi(\tau)&=&\phi_{R}(\tau)+\phi_{L}(\tau),
\end{eqnarray}
where
\begin{subequations}
\begin{align}
\phi_{R}(\tau)&=\!\!\!\int d\Omega N_{\frac{\Omega}{a}}e^{-\frac{1}{2}(\tilde{\lambda}-\Omega)^{2}}\text{ch}_{\Omega}e^{-i\Omega\tau}a_{\Omega,R}+\text{h.c.}\, ,\\
\phi_{L}(\tau)&=\!\!\!\int d\Omega N_{\frac{\Omega}{a}}e^{-\frac{1}{2}(\tilde{\lambda}-\Omega)^{2}}\text{sh}_{\Omega}e^{+i\Omega\tau}a_{\Omega,L}+\text{h.c.}\, ,
\end{align}
\end{subequations}
and to shorten notation we have defined $\text{ch}_{\Omega}:=\cosh(r_{\Omega/a})$ and $\text{sh}_{\Omega}:=\sinh(r_{\Omega/a})$.
Note that the effective interaction strength is now modulated by hyperbolic trigonometric functions. We are currently using this detector model to analyse entanglement extraction of sharp frequency Unruh states. As uniform acceleration is also a stationary orbit of flat spacetime, we expect a time independent vacuum transition rate~\cite{letlaw1981}. Using the parametrised Unruh modes, we can calculate the transition rate of the accelerated detector. We find for the field in its vacuum state,
\begin{eqnarray}
\label{acceleratedresponse}
\dot{F}_{\tau}(\Delta)&=&\frac{1}{e^{2\pi\Delta/a}-1}\Xi\left(\Delta\right),
\end{eqnarray}
where
\begin{eqnarray}
\begin{array}{ccl} 
\Xi\left(\Delta\right)&:=&\int d^{2}\vec{k}_{\perp}\left[N^{2}_{\vec{\Delta}}|\tilde{f}(\vec{\Delta})|^{2}\Theta(\Delta)\right.\\
& &\hspace{7mm}-\left. N^{2}_{-\vec{\Delta}}|\tilde{f}(-\vec{\Delta})|^{2}\Theta(-\Delta)\right],
\end{array}
\end{eqnarray}
with $\pm\vec{\Delta}:=(\pm\Delta,k_{y},k_{z})$ and $N_{\vec{\Delta}}$ denotes the appropriate normalisation for the Rindler modes. We can see immediately the transition rate of the detector is the expected thermal distribution, where the temperature is inversely proportional to the acceleration parameter $a$, but again modified by the frequency of the field operator. Again as examples, we plot the transition rate~(\ref{acceleratedresponse}) for the $(1+1)$ massless field for different accelerations. In FIG.~(\ref{fig:1+1accelerateddelta}), we plot the transition rate for a Dirac delta profile. In FIG.~(\ref{fig:1+1acceleratedgaussianeps}), we plot transition rate for a double Gaussian speaking profile. We note the qualitative and quantitative differences between the standard point-like detector and our modified model. 

We also note that Eq.~\eqref{acceleratedresponse} satisfies the Kubo-Martin-Schwinger condition \cite{hodgkinson2012-1,hodgkinson2012-2}
\begin{eqnarray}
\dot{F}_{\tau}(\Delta)=e^{-\frac{2\pi}{a}\Delta}\dot{F}_{\tau}(-\Delta).
\end{eqnarray}
The transition rate is, as expected, independent of time due to the stationarity of the trajectory and the invariance of the vacuum state. Now we shall analyse the response of our accelerated detector model when the field contains a single Unruh particle. In the literature, well analysed states of the field correspond to maximally entangled Bell states, see for example, \cite{alsing2003,friis2011,datta2009}. These states contain both the vacuum and a single Unruh particle. Our starting point will again be the Wightman-function which, for the one particle state, takes the form
\begin{eqnarray}
W(\tau,\tau')&:=&\bra{1_{p}}\phi(\tau)\phi(\tau')\ket{1_{p}},
\end{eqnarray}
where $\ket{1_{p}}$ is a one Unruh particle state defined as $\ket{1_{p}}:=\int d\Omega  d^{2}\vec{k}_{\perp} \Phi(\Omega,\vec{k}_{\perp})a^{\dag}_{\Omega,\vec{k}_{\perp},p}\ket{0}$. Continuing in the exact same fashion as the Minkowski one particle state we find 
\begin{equation}
\begin{aligned}
\label{Eq:wightmann1paccelerated}
W(\tau,\tau')&=\int d\Omega d^{2}\vec{k}_{\perp} |\Phi(\vec{k})|^{2}\cdot \bra{0}\phi(\tau)\phi(\tau')\ket{0}\\
& \hspace{5mm}+2\text{Re}\left[I_{p}^{*}(\tau)I_{p}(\tau')\right],
\end{aligned}
\end{equation}
where $I_{p}(\tau):=\int d\Omega d^{2}\vec{k}_{\perp}\tilde{f}(\Omega,\vec{k}_{\perp})\Phi(\Omega,\vec{k}_{\perp})U_{\Omega,p}(\tau)$ with
\begin{eqnarray}
   U_{\Omega,p}(\tau) = N_{\Omega/a}\left\{
     \begin{array}{rcl}
        \cosh(r_{\Omega/a})e^{-i\Omega\tau} & : & p=R\\
        \sinh(r_{\Omega/a})e^{+i\Omega\tau} & : & p=L
     \end{array}
   \right. .
\end{eqnarray}
An Unruh particle has with it an associated wavefunction, defined as $\int d\Omega d^{2}\vec{k}_{\perp} \Phi(\Omega,\vec{k}_{\perp})U_{\Omega,p}(t,x)$. The accelerated detector will probe the Unruh particles wavefunction as it approaches. As the particles wavefunction oscillates as a function of $\tau$, the detector will observe different phases at different times. It is these oscillations that contribute to the undulatory behaviour of the detectors transition rate. For the Unruh state, the corresponding accelerated expression for Eq.~(\ref{iota}) again has a time dependent oscillatory integral. It is clear, for appropriately behaved functions $\tilde{f}$ and $g$, the same analysis can be applied here as for the Minkwoski particle. The Riemann-Lebesgue lemma can be used to show in the asymptotic past and future, the response of the detector is the same as the vacuum and hence has a thermal signature. As with the Minkowski particle, the intermediate regions between past and future asymptotic times give rise to an oscillatory response function. In the limit of high acceleration, the second term in Eq.~\eqref{Eq:wightmann1paccelerated} becomes negligible and the state tends to the maximally mixed state. Thus, the single particle state of the field is dominated by the vacuum fluctuations.

\section{Conclusions}
We introduce a detector model which naturally couples to peaked frequency distributions of Minkowski, Unruh and Rindler modes.  This detector model is suitable for studies of entanglement extraction in non-inertial frames. In the $(3+1)$-dimensional case, the frequency window of the detector peaks around positive and negative momentum inducing a double peaking. In the (1+1)-dimensional case, frequency distributions naturally peak around a single frequency. We obtain analytical results for the instantaneous transition rates of the detectors undergoing inertial and uniformly accelerated motion. In particular, the transition rate of the accelerated detector is the expected thermal distribution modified by a smearing function that arises from the detectors spatial profile. We have also shown the well studied single Unruh particle states produce an oscillatory response that is only thermal in the asymptotic past and future. Since for accelerated detectors thermal noise is observed in both the vacuum and the single Unruh particle state, entanglement is expected to be degraded by the Unruh effect for global Unruh modes. We also see the degradation effects occur for properly normalised wave packets.

As shown in figures FIGS~[2-5], the response of the Unruh-DeWitt detector depends strongly on the profile of its interaction and, thus, on the physical system that implements the detector. Contributing factors to this profile will be the eigenstates of the detector, external components such as driving laser fields and the geometry of the system. We have shown the response of the detector can vary significantly and this will have observable consequences in the experimental verification of results in quantum field theory and relativistic quantum information. To this end, we comment on the physical feasibility and impact on constructing the spatially dependent coupling profile we have described.

A spatially dependent coupling strength can be engineered by placing the quantum system in an external potential which is time and space dependent. These tuneable interactions have been produced in ion traps~\cite{thompson1990,miller2005}, cavity QED~\cite{walther1006} and superconducting circuits~\cite{peropadre2010,gambetta2011,srinivasan2011,sabin2012}. In an ion trap, the interaction of the ion with its vibrational modes can be modulated by a time and spatial dependent classical driving field, such as a laser~\cite{haffner2008}. In cavity QED, time and space dependent coupling strengths are used to engineer an effective coupling between two cavity modes~\cite{imamoglu1997,guzman2006}. Within this setting, our work is particularly relevant where current investigations ~\cite{chen1999,scully2003,hu2004} study the effect of artificial atoms, with large spatial profiles, interacting with modes of an electromagnetic field. Our model can also be, in principle, realized in a Bose-Einstein condensate where a harmonic oscillator detector corresponds to an impurity created by a potential well within the condensate~\cite{colombe2007,shapiro2009,kumar2012}. The spatial profile is engineered by choosing the right shape of the trapping potential. The harmonic oscillator couples to the phononic field of the condensate, which obeys a Klein-Gordon equation in an effective curved metric. Therefore, our model can be used to study effects in analogue gravity.


Unruh-DeWitt type detectors have allowed us to explore different coupling scenarios in quantum field theory. In tandem, our results have paved the way for the development of novel mathematical techniques that allow a non-perturbative treatment of Unruh-DeWitt detectors~\cite{bruschi2013-2,brown2013} and more advanced phenomena, such as detector-field back reaction~\cite{Lin2007}. With these considerations, we hope that Unruh-DeWitt detectors can be used to gain further insight into quantum information within quantum field theory.\\ \\
The authors would like to thank N. Friis, D. E. Bruschi, J. Louko, A. Dragan, L. Hodgkinson, G. Adesso, S. Tavares, S.Y. Lin and T. Tufarelli for useful comments and discussions. I. F. was supported by EPSRC [CAF Grant EP/G00496X/2].

\bibliography{references}

\end{document}